\documentclass[]{interact}

\usepackage[utf8]{inputenc}
\usepackage{placeins}
\usepackage[noend]{algpseudocode}
\usepackage{algorithm}
\usepackage{algorithmicx}
\usepackage[dvipsnames]{xcolor}
\usepackage{makecell}
\graphicspath{ {./images/} }

\usepackage{epstopdf}
\usepackage[caption=false]{subfig}

\usepackage{natbib}
\bibpunct[, ]{(}{)}{;}{a}{}{,}

\theoremstyle{plain}

\theoremstyle{definition}

\theoremstyle{remark}

\algnewcommand\algorithmicinput{\textbf{Input:}}
\algnewcommand\Input{\item[\algorithmicinput]}

\algnewcommand\algorithmicparameters{\textbf{Parameters:}}
\algnewcommand\Parameters{\item[\algorithmicparameters]}

\algnewcommand\algorithmicstored{\textbf{Stored state:}}
\algnewcommand\Stored{\item[\algorithmicstored]}

\algnewcommand\algorithmicmyreturn{\textbf{Return:}}
\algnewcommand\MyReturn{\item[\algorithmicmyreturn]}

\begin{document}

\articletype{ARTICLE}

\title{Fault detection in propulsion motors in the presence of concept drift}

\author{
  \name{Martin Tveten\textsuperscript{a}\thanks{CONTACT Martin Tveten. Email: tveten@nr.no} and Morten Stakkeland\textsuperscript{b}}
  \affil{\textsuperscript{a}Norsk Regnesentral, Gaustadalleen 23a, Kristen Nygaards hus, 0373 Oslo, Norway\\ \textsuperscript{b}ABB, PO. Box 129, 1325 Lysaker, Norway}
}

\maketitle
\begin{abstract}
Machine learning and statistical methods can improve conventional motor protection systems, providing early warning and detection of emerging failures. Data-driven methods rely on historical data to learn how the system is expected to behave under normal circumstances. An unexpected change in the underlying system may cause a change in the statistical properties of the data, and by this alter the performance of the fault detection algorithm in terms of time to detection and false alarms. This kind of change, called \textit{concept drift}, requires adaptations to maintain constant performance. In this article, we present a machine learning approach for detecting overheating in the stator windings of marine electrical propulsion motors.  Using simulated overheating faults injected into operational data, the methods are shown to provide early detection compared to conventional methods based on temperature readings and fixed limits. The proposed monitors are designed to operate for a type of concept drift observed in operational data collected from a specific class of motors in a fleet of ships. Using a mix of real and simulated concept drifts, it is shown that the proposed monitors are able to provide early detections during and after concept drifts, without the need for full model retraining.
\end{abstract}


\section{Introduction}
In this article, we study detection of overheating in the stator winding temperatures of marine electric propulsion motors. Traditionally, motor protection on this kind of failure modes is implemented by installing temperature sensors on the windings, and setting a fixed alarm limit where the motor is shut down \citep{ieee_ieee_2017}. Given the physical size and mass of megawatt rated propulsion motors, there is a risk that heat developing in a hotspot avoids detection by the installed sensors before a failure occurs. 
Taking into account the criticality of the motors, we propose adding monitoring to reliably provide early warnings in case of emerging failures. 

Several authors explore different approaches for detecting stator winding faults and motor overheating.
\citet{cipriani_automatic_2021} and \citet{xu_deep_2024} use thermal sensors, for instance a Forward Looking Infrared (FLIR) camera, and image detection techniques for detecting abnormal temperatures.
Retrofitting thermal monitoring systems onto existing propulsion motors may be theoretically possible, but very challenging or in practice.
Another approach is to use high frequency multi-phase current measurements and perform motor current signature analysis (MCSA).
In \citet{otava_implementation_2016}, winding short circuits and resulting phase imbalance in a permanent magnet synchronous motor are detected using millisecond sampling of the current phases. Similarly, winding faults in both induction and synchronous motors are detected by estimating the Park's vector in \citet{cruz_stator_2001}.
Methods based on motor current signature analysis do not catch other potential causes of overheating, such as over-powering or cooling system malfunctioning.
Moreover, gathering operational data from ships sampled at 1Hz over longer periods of time introduces a practical data management challenge. Lastly, it can be argued that catching a short circuit in the windings after it has occurred is too late, the motor is already damaged. 


Recently, data-driven approaches are proposed and applied for condition monitoring \citep{karatug_review_2023}, fault diagnosis \citep{youssef_survey_2024}, and fault detection \citep{velasco-gallego_recent_2023}. In this paper we explore the use of statistical and machine learning methods for fault detection.  
More specifically, we explore and develop methods based on \textit{temperature residuals}.
This means that we monitor the difference between actual and expected motor temperatures.
The expected temperatures are generated from a model of the temperatures given operational characteristics of the motor.
As outlined in \citet{wallscheid_thermal_2021}, common approaches to modelling expected motor temperatures range from physical modelling to purely data-driven models.
For an example of a physical model of the temperature of the stator windings of a permanent magnet synchronous motor see \citet{guo_real-time_2023}.
On the other side of the spectrum, \citet{wallscheid_investigation_2017} uses a long short-term memory deep learning model.
In \citet{kirchgassner_empirical_2019}, input variables are smoothed using exponentially weighted moving average (EWMA) filters and combined in a linear model to model several motor temperatures on a permanent magnet synchronous motor, including stator winding temperatures. 
Similarly, EWMA filters and a linear model was used to model stator winding temperatures in a synchronous, brushless marine propulsion motor in  \citet{hellton_real-time_2021}. 
Using a changepoint detection algorithm based on cumulative sums (CUSUM), it was shown in \citet{hellton_real-time_2021} that a specific fault incident could have been detected two hours prior to the actual shutdown of the system. 

\subsection{Concept drift}
\label{sec:introconcept}
Machine learning approaches to overheating detection or fault detection in general comes with their own set of challenges, however.
The focus of this article is how to deal with unforeseen changes in the data behaviour, known as \textit{concept drift}.
See \citet{lu_learning_2019} and \citet{lima_learning_2022} for a review and general introduction to the topic.
Changing behaviour over time is a challenge for machine learning models because they are trained based on prior data.
If the training data is not representative of the current system behaviour, fault detection performance will be poor.
Dealing with concept drift is therefore essential to ensure that the fault detection algorithm is reliable over time.

We present data from several ships with real cases of concept drift in Section \ref{sec:real_drift_description}.

\begin{figure}[tb]
    \centering
    \includegraphics[width=0.9\textwidth]{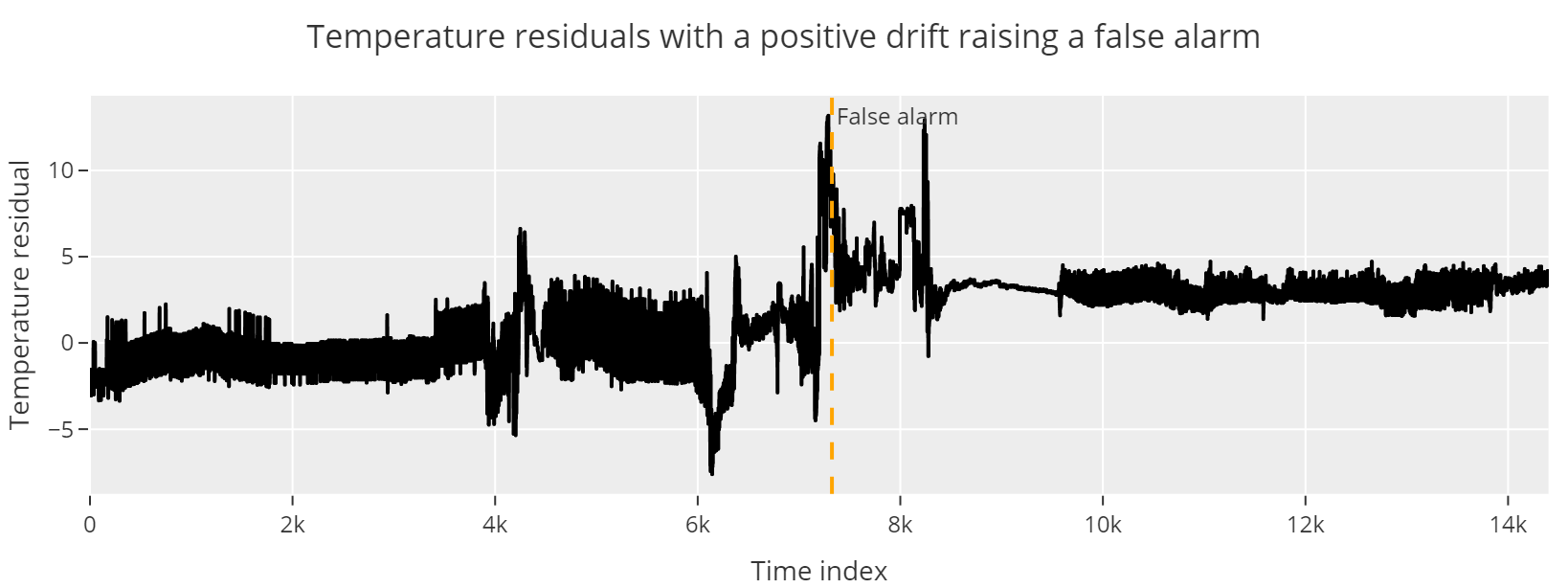}
    \includegraphics[width=0.9\textwidth]{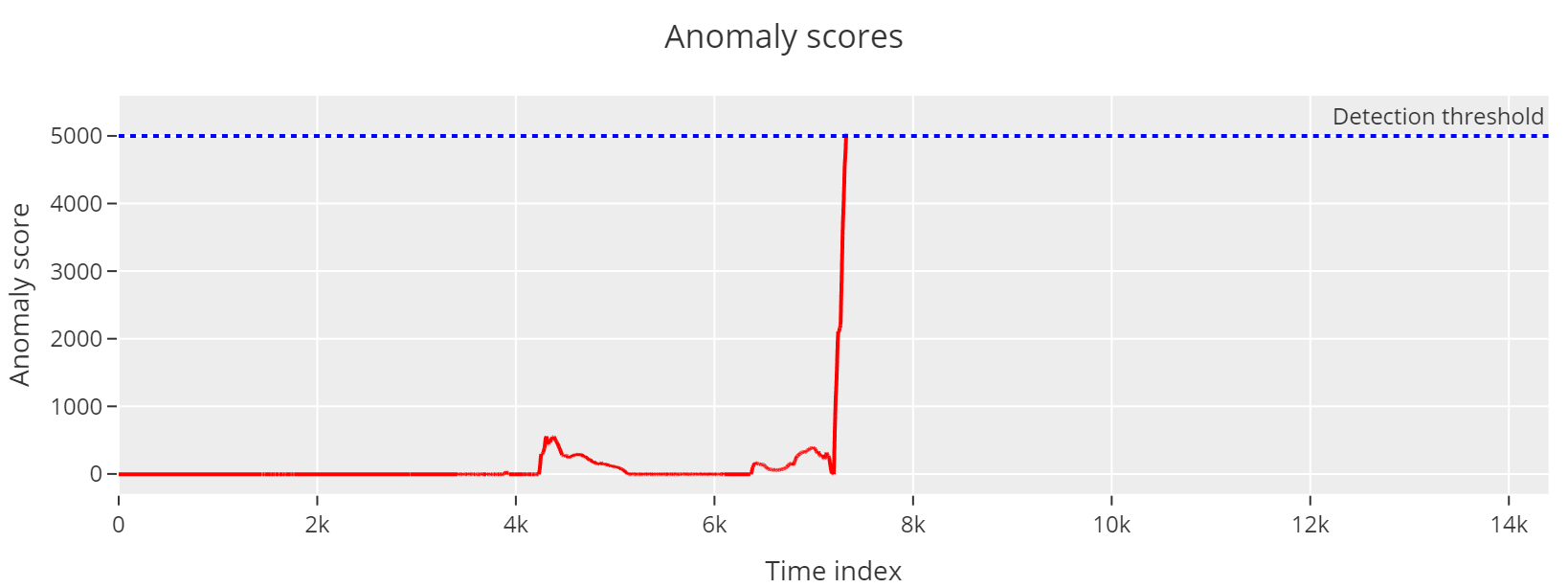}
    \caption{A rise in baseline temperature leads to a rise in the residuals which again triggers a false alarm.}
    \label{fig:temp_rise}
\end{figure}
\begin{figure}[tb]
    \centering
    \includegraphics[width=0.9\textwidth]{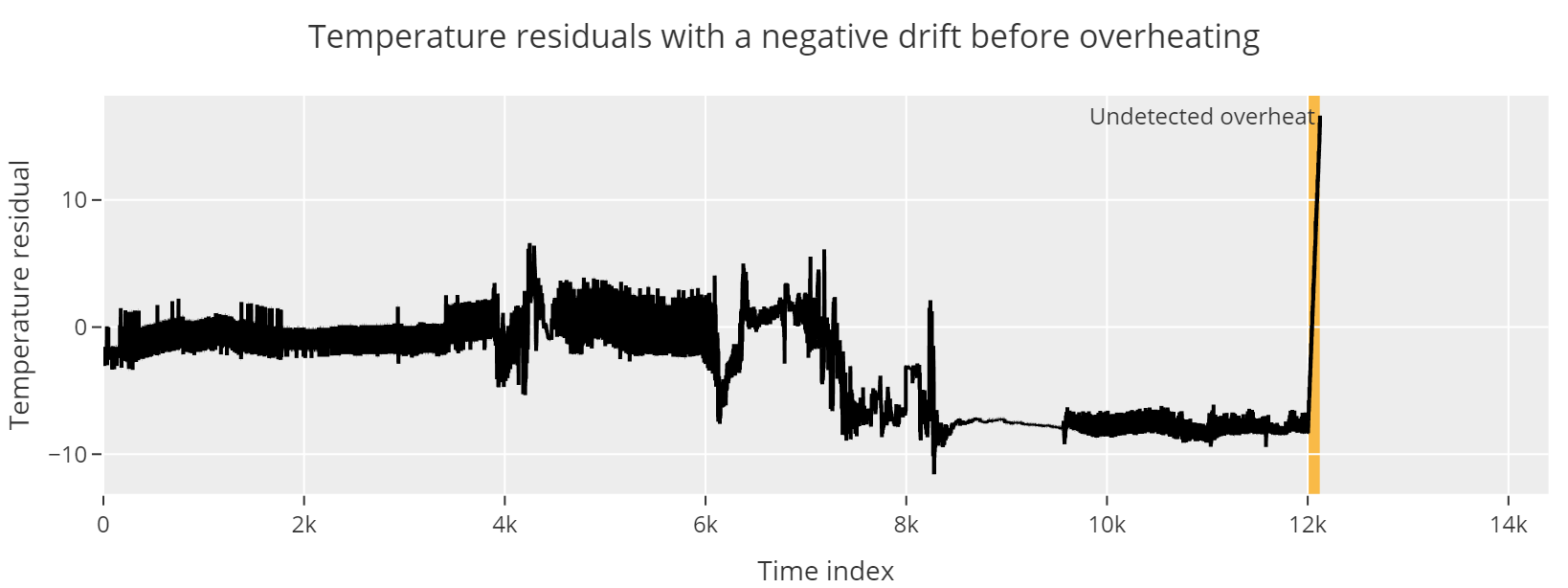}
    \includegraphics[width=0.9\textwidth]{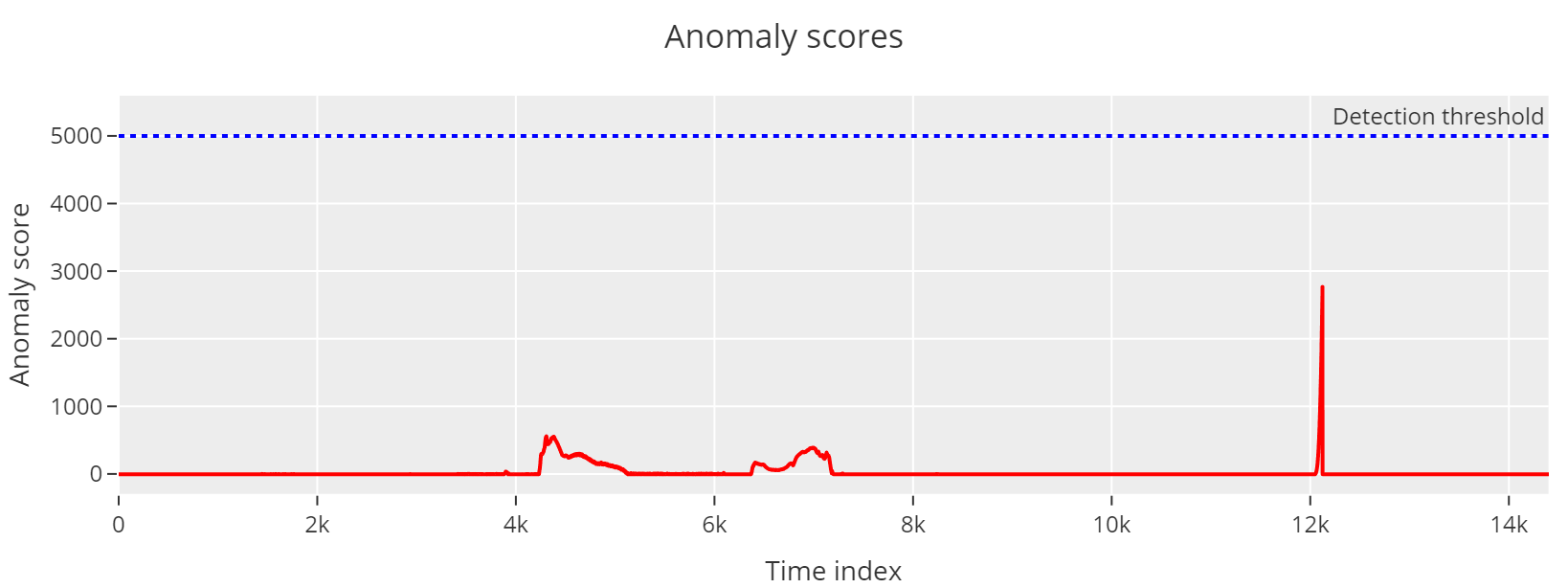}
    \caption{The baseline temperature decreases, leading to a negative bias in the residuals, masking an overheating event.}
    \label{fig:temp_drop}
\end{figure}

To illustrate the effected of untreated concept drift in a fault detection system, we consider two examples.
In both examples, a machine learning model for predicting the motor temperature is trained on data before a concept drift occurs, and the temperature residuals are monitored to detect faults.
\begin{itemize}
    \item \textbf{Example 1: Increased risk of false alarms.}
    Consider a case where the fans in the cooling system are replaced with a new type with less effective cooling.
    Figure \ref{fig:temp_rise} illustrates what happens in this scenario. The consequence is that the baseline temperature increases after the fans are replaced (time index 7100).
    In other words, the model starts to systematically underestimate the true temperature, giving larger positive residuals than before. 
    This increases the risk of false alarms.
    \item \textbf{Example 2: Blinding.} Now consider the opposite case, where new fans in the cooling system yield better cooling than before.
    An illustration is given in Figure \ref{fig:temp_drop}.
    In this case, the model systematically overestimates the true temperature, giving larger negative residuals than before.
    This delays the time to detection and increases the risk of missing the overheating event, potentially "blinding" the fault detection algorithm.
\end{itemize}

We conclude this section by a review of strategies to deal with concept drift.
The strategies can broadly be put in three different categories:
on-demand adaptation, continuous adaptation and periodic retraining.

The first strategy is to continuously monitor the performance of the prediction model, and retrain the model as soon as performance drops consistently by a sufficient amount.
That is, a drift is adapted to \textit{on demand} by first detecting that a drift has occurred, then adapting to it.
This is the most common strategy encountered in the concept drift literature.

A popular method is the Adaptive Windowing (ADWIN) method of \citet{bifet_learning_2007}, where the means of two sliding, adaptive windows of the residuals are compared.
Large differences in the mean result in drifts being detected.
Similarly, the Just-In-Time (JIT) method of \citet{alippi_just--time_2008} use a CUmulative SUM (CUSUM) based approach to detect changes in the mean.
These methods use the same conceptual idea as the method we propose in Section \ref{sec:abrupt_concept_drift}, but they don't provide sufficient customisation of the windows to work well in an anomaly detection context.
See \citet{ma_robust_2018} or \citet{jourdan_handling_2023} for examples in relation to condition monitoring.



A second strategy is \textit{continuous adaptation}, where the prediction model is continuously updated for every new sample \citep{grote-ramm_continual_2023}.
This is also known as \textit{online learning} in the machine learning literature \citep{hoi_online_2021}.
Such a strategy implicitly assumes that drifting behaviour occurs incrementally all the time and skips the step of detecting whether or not a drift has occurred.

An example of such a strategy is to use a linear regression model, and update its parameters by an incremental stochastic gradient descent algorithm \citep{ram_stochastic_2007}.

Lastly, for completeness, we also describe the \textit{periodic retraining} approach. Periodic retraining means to retrain the entire model at periodic intervals, say every week or every month.
This is arguably the most common approach in practice, but it is not an acknowledged approach in the scientific literature on concept drift due to its ad-hoc nature.
A fundamental issue with periodic retraining occurs if the model is retrained too rarely.
In that case, a drift may persist untreated in the data for a sufficient amount of time for the false alarm rate and time to detection to be severely affected. 
Setting a high retrain rate using all available historical data would add a major computational burden on the system. The computational load can be reduced by limiting the training data to within a sliding window of fixed duration, but the window needs to be long enough to train a model with sufficient accuracy. In addition, the accuracy of the model and the performance of the fault detector will be unpredictable in the cases where the window contains data from both before and after a concept drift.   

\subsection{Our contributions}
In this article we propose a general methodology for dealing with concept drift in real-time fault detection methods based on monitoring residuals from machine learning prediction models.
The methodology is based on introducing a drift adjustment parameter in such a way that retraining the entire machine learning model is avoided.
We also present two concrete methods for adapting the model to concept drift within our framework.
The first method is an on-demand adaptor, while the second method is a continuous adaptor.

Finally, the methods are applied for detecting overheating in the stator windings in electrical propulsion motors.
The performance of the methods is evaluated on a large amount of real data from operational ship motors, where we insert simulated overheating events and use a mix of real and simulated concept drifts.
The methods are shown to increase fault detection performance for drifts that occur as a sudden shift in the mean of the data.


\section{Data} \label{sec: data}
\subsection{System}
\begin{figure}[tb]
    \centering
    \includegraphics[width=1\linewidth]{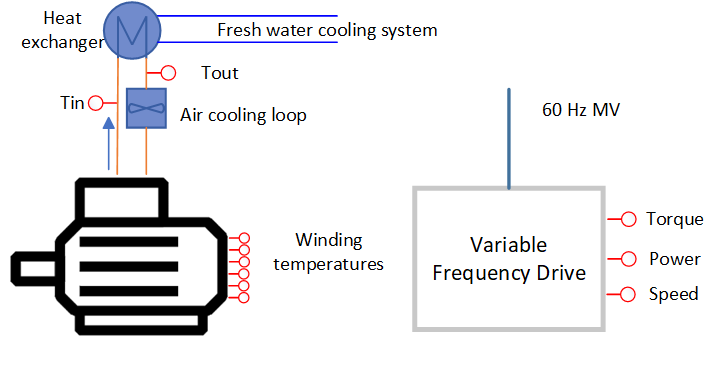}
    \caption{System overview of an electrical propulsion motor with a Variable Frequency Drive.}
    \label{fig:system_overview}
\end{figure}

The systems we consider are electrical propulsion motors in a diesel-electric power distribution system. The data is collected from synchronous, double-winded electrical propulsion motors, with brushless excitation of the rotor. A system overview can be seen in Figure \ref{fig:system_overview}. The motor is air cooled, where the excess heat in the air cooling loop is removed by a heat exchanger.  Based on the input from the bridge a propulsion control unit sets the required torque and speed. The motors are controlled by a frequency converter, often called a variable speed drive. The drive also provides protection for the propulsion motors and collects relevant sensor readings from the cooling system.

\subsection{Dataset} \label{sec:dataset}
Data is collected from fifteen ships with three motors each, giving a total of 45 propulsion motors.
The amount of data from each motor varies, ranging from approximately one to five years.
In total there are 162 years of measurements.
All measurements are pre-processed as minute-wise averages. For each motor, the following variables are used for temperature modelling: 
Six stator winding temperatures, cooling air inlet temperature,  power, speed, and torque. 
In addition, a measurement of the cooling water temperature internally in the variable speed drive is used as an explanatory variable. As no measurement of the temperature of the onboard fresh water cooling system is available, this functions as a proxy measurement.

\subsection{Concept drift in the motor data} \label{sec:real_drift_description}
\begin{figure}[tb]
    \centering
    \includegraphics[width=0.9\textwidth]{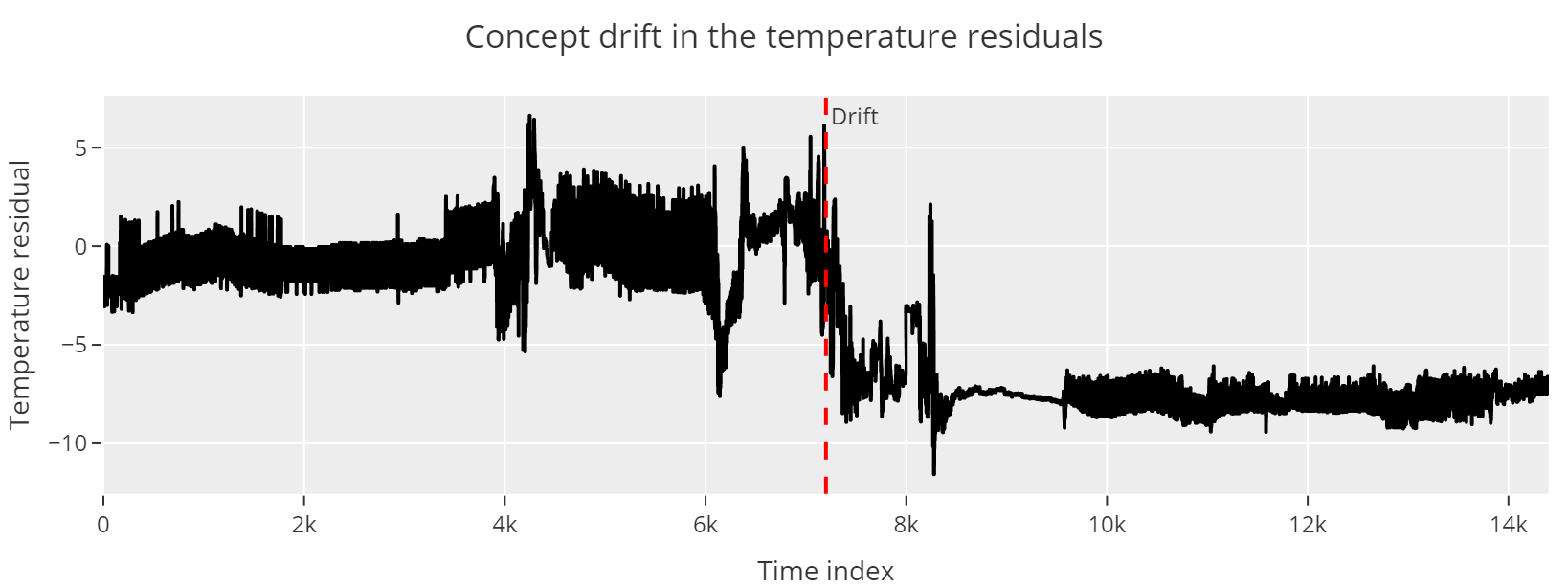}
    \caption{A real case of concept drift observed in one of the residual temperature series. After the drift, a persistent offset of around $-7$ appears in the residuals.}
    \label{fig:real_drift}
\end{figure}

During a project aimed to productise a fault detector based on \citet{hellton_real-time_2021}, we discovered a discrete shift in the mean of the measured temperatures for several ships.
An example of an observed change in the temperature residuals is shown in Figure \ref{fig:real_drift}. 
Before the drift, the values are centred around zero, as expected, while after the drift, the values are centred around -7. The cause of the changes was later identified as a maintenance operation which modified the cooling regime in the motors.

\section{Methods}\label{sec:methods}
\subsection{Real-time fault detection} \label{sec:fault_detection}

To develop the fault detection algorithm, we follow the two-step approach of \citet{hellton_real-time_2021}:
\begin{enumerate}
    \item Use historical data to train a machine learning model for predicting the winding temperatures from the remaining variables.
    \item Monitor the temperature residuals of observed versus predicted winding temperature on live data. 
    When the residuals become sufficiently large for a long enough time, an alarm is raised.
\end{enumerate}

For the fault detection algorithm to be useful, it is is essential to keep a strict control on the number of false alarms. More false alarms means a less trustworthy monitoring system and implies a high risk of alarms being ignored.
Conversely, detecting the onset of an actual fault earlier provides more time to prevent catastrophic outcomes.
We therefore have the following two performance requirements:
\begin{enumerate}
    \item Minimize false alarms: The algorithm should generate as few false alarms as possible, and at least under some acceptable level.
    \item Early fault detection: The algorithm should identify the beginning of actual faults as soon as possible, and at least before the trip limit is reached.
\end{enumerate}
From a statistical perspective, the first requirement is related to minimizing the number of false positives, while the second requirement is related to minimizing false negatives.

Our scope with respect to simultaneous fault and drift detection is given as follows.
\begin{enumerate}
    \item Faults are characterized by large, positive change in the mean of the temperature residuals over a short time period, on the order of a few hours. 
    \item Drifts are also characterised by changes in the mean of the temperature residuals, but the changes are of smaller magnitude than faults, and it can change in both a positive and negative direction. In addition, drifts last for a much longer time than faults; a change is not a drift unless it lasts for more than a day. Also note that we are primarily interested in sudden drifts. 
    \item Known faults are extremely rare. That is, we cannot train a classifier on normal versus faulty data. 
    \item We are concerned with safety-critical applications, in the sense that both missed detections and false alarms are severe and should be at a very low level. 
\end{enumerate}
Note that distinguishing drifts and faults by their duration has been done in similar problems before, see for example \citet{liu_anomaly_2023}.

\subsection{High-level algorithm}
We now describe our general fault detection methodology with concept drift adaptation in mathematical terms.
A high-level description of the algorithm is summarised in Algorithm \ref{alg:high-level-alg}.
Let $y_{j, t}$ for $j = 1, \ldots, p$ be the winding temperature for sensor $j$ at time $t$.
The vector of all $p$ winding temperatures at $t$ is denoted by ${y_t = (y_{1, t}, \ldots, y_{p, t})}$. 
$x_t$ denotes the vector of input variables to the prediction model at each time $t$.
The first step is to train a prediction model $\hat{f}$ on historical data for generating temperature predictions $\hat{y}_t = \hat{f}(x_t)$.
When running on live data the temperature residuals $e_t = y_t - \hat{y}_t$ are monitored over time.
To account for possible concept drift, the residuals are fed to an $\text{AdaptDrift}$ function that returns \textit{adapted temperature residuals} $\tilde{e}_t$.
Finally, $\text{DetectAnomaly}$ performs anomaly detection on the adapted residuals, returning $A_t = 1$ if an alarm is raised at time $t$ and $A_t = 0$ otherwise.

\begin{algorithm}[H]
  \caption{High-level algorithm for fault detection with drift adaptation}
  \label{alg:high-level-alg}
  \begin{algorithmic}[1]
   \Input Pre-trained prediction model $\hat{f}$.
   \For{$t = 1, 2, \ldots, $}
        \State Get temperatures $y_t$ and input variables $x_t$
        \State$\hat{y}_t = \hat{f}(x_t)$
        \State$e_t = y_t - \hat{y}_t$
        \State$\tilde{e}_t = \text{AdaptDrift}(e_t)$
        \State$A_t = \text{DetectAnomaly}(\tilde{e}_t)$ \Comment{$A_t$ is 1 if an alarm is raised, 0 otherwise.}
    \EndFor
    \MyReturn $\mathcal{A} = \{t: A_t = 1\}$ \Comment{The alarm times.}
  \end{algorithmic}
\end{algorithm}

Our general solution to adapting to concept drift is to use an additive \textit{drift adjustment} $\hat{b}_t$, where $\hat{b}_t$ is estimated in an online manner.
That is, we take
\begin{equation}
    \text{AdaptDrift}(e_t) = e_t + \hat{b}_t,
\label{eq:general_drift_adaptor}
\end{equation}
in Algorithm \ref{alg:high-level-alg}.
Implicitly, this means that we allow for the intercept term in the prediction model to be time-varying.

The reason for this choice of drift adaptor is that only a single parameter must be updated, rather than the entire prediction model.
Moreover, it is agnostic to the choice of prediction model; it works regardless of the model being a neural network, a boosted tree model or a simple linear regression.
The results (Section \ref{sec:results}) indicate that such a drift adaptation method is adequate in the case of overheating detection. 

Next we give a detailed description of choices and parts of Algorithm \ref{alg:high-level-alg}.
First, our temperature prediction model is explained in Section \ref{sec:prediction_model}.
We then describe two drift adaptation methods in Section \ref{sec:concept_drift_adaptation},
before the anomaly detection method is detailed in Section \ref{sec:anomaly_detection_method}

\subsection{Temperature prediction} \label{sec:prediction_model}

As our prediction model for the motor temperature, we use a gradient boosted decision tree model.
Such models are originally due to \citet{friedman_greedy_2001}. 
For an introduction to the topic, see \citet{hastie_elements_2009}.
Gradient boosted decision tree models are an ensemble learning technique that build a series of decision trees in a sequential manner.
Each tree is designed to correct the errors of its predecessor, ultimately combining their outputs for improved prediction accuracy.
We use this model as it supports different data types and handles non-linear relationships between variables without extensive preprocessing.
In addition, it is known to offer competitive prediction accuracy across a range of different regression problems \citep{chen_xgboost_2016}.

To predict the motor temperatures, we train one gradient boosted decision tree model with squared error loss per motor.
The HistGradientBoostingRegressor of Scikit-learn \citep{pedregosa_scikit-learn_2011} is used to fit the models due to its ease of use.
We use five-fold cross-validation to tune the three hyperparameters maximum tree depth, the maximum number of iterations and the learning rate of the model.
The tree depth is the number of edges to go from the root to the deepest leaf of each decision tree, 
the number of iterations are the number of trees in the model,
and the learning rate scales the contribution of each new tree that is added to the model.
The defaults are used for the remaining hyperparameters.

Following \citet{kirchgassner_empirical_2019}, and \citet{hellton_real-time_2021}, we use EWMA smoothing for some of the input variables. Given a sequence of inputs $x_k$, the EWMA smoothed $x$ is given by

\begin{equation}
   e_k = (1 - \alpha) x_k + \alpha e_{k-1}
\end{equation}
The half time $\tau$ of the EWMA process is defined as the number of steps required for an input value to reach half of its original value.
This relationship is given by $\alpha=(1/2)^{1/\tau}$.

The full list of input features is:
\begin{enumerate}
    \item Power and EWMA-transformed power with 30 minutes half-life.
    \item Speed and EWMA-transformed speed with 30 minutes half-life.
    \item Torque and EWMA-transformed torque with 30 minutes half-life.
    \item Air inlet cooling temperature.
    \item Water cooling unit temperature.
    \item The time since the motor was switched on or off.
    \item Duration of previous on/off state.
    \item A categorical variable indicating sensor number.
\end{enumerate}

Note that an on/off state is added and used to derive two input variables. The reason for adding these derived times is to provide the model information that can be used to learn the burn-in dynamics after the switching on or off.
The categorical variable per sensor effectively allows the model to adapt to differences between the sensors. Due to different mounting locations, the temperature readings differ under the same operating conditions.

\subsection{Concept drift adaptation} \label{sec:concept_drift_adaptation}

\subsubsection{On-demand drift adaptation} \label{sec:abrupt_concept_drift}
For on-demand drift adaptation of sudden concept drifts, we first run a changepoint test to test for the presence of a changepoint in the residuals.
If a change is detected, then the drift adjustment $\hat{b}_t$ in \eqref{eq:general_drift_adaptor} is retrained as the mean of the residuals over a period following the detection time.

A good candidate for detecting such drifts is the CUSUM statistic which was first described by ES Page \citep{page_continuous_1954}. This metric aims to detect a change in mean of a variable, that occurs at a given changepoint.
For a single sensor $j$, changepoint candidate $s$ and current time $t$, the CUSUM statistic is defined as
\begin{equation}
    \mathcal{C}_{j, s+1:t} = S_{j, s+1:t} /\sqrt{t-s},
    \label{eq:drift-cusum}
\end{equation}
where $S_{j, s+1:t} = \sum_{i=s+1}^t e_{j, i}$.
The drift score for sensor $j$ is obtained by taking the maximum over a selected set of candidate changepoints ${\mathcal{S} \subseteq \{1, \ldots, t-1 \}}$:
\begin{equation}
    \mathcal{C}_{j, t} =  \max_{s \in \mathcal{S}} |\mathcal{C}_{j, s+1:t}|.
    \label{eq:drift-score}
\end{equation}
To get some intuition for the drift score, observe that the CUSUM statistic \eqref{eq:drift-cusum} is equal to $\sqrt{t-s}\bar{e}_{s+1:t}$, where $\bar{e}_{s+1:t} = S_{s+1:t}/(t-s)$ is the mean of the residuals since a candidate changepoint.
This means that the CUSUM statistic can be large not only if the mean is large but the period $(s, t]$ is relatively short, but also if the mean is relatively small but the period is long.
It reflects the idea of detecting a concept drift if it is 'sufficiently large for a sufficiently long time', where there is an explicit trade-off between the size of the change in mean and the length of the period it is calculated over.

Importantly, notice that the set of candidate changepoints $\mathcal{S}$ is fully customisable.
This allows us to only consider changepoints that occur sufficiently far back in time as drifts.

To get a test over all sensors, we aggregate the individual drift scores by summing them.
A drift is detected as soon as the summed scores raises above a threshold $\lambda$.
This gives us the global drift score
\begin{equation}
    \mathcal{C}_t = \sum_{j=1}^p \mathcal{C}_{j, t} > \lambda,
    \label{eq:drift-test}
\end{equation}

Using this changepoint test statistic as our drift detector is a suitable choice due to the characteristics of the drifts we are interested in.
Firstly, it uses the information that the mean of the residuals should be $0$ if the prediction model is unbiased.
Secondly, it can detect both positive and negative changes in the mean of the residuals.
And thirdly, it allows us to customise the candidate changepoints such that only sufficiently long periods to count as a drift may be considered.

A final feature we incorporate in the drift detector is a lag parameter $l$ in a lagged version of the CUSUM statistic, defined by
\begin{equation}
    \mathcal{C}_t(l) = \mathcal{C}_{t-l}.
\end{equation}
The lag parameter's role is to avoid including observations during periods of real faults, which, in a worst-case scenario, could lead to the concept drift adaptor masking a real fault.
The lag $l$ should therefore ideally be set to be larger than the longest feasible duration of an overheating event.

\begin{figure}[tb]
    \centering
    \includegraphics[width=0.9\textwidth]{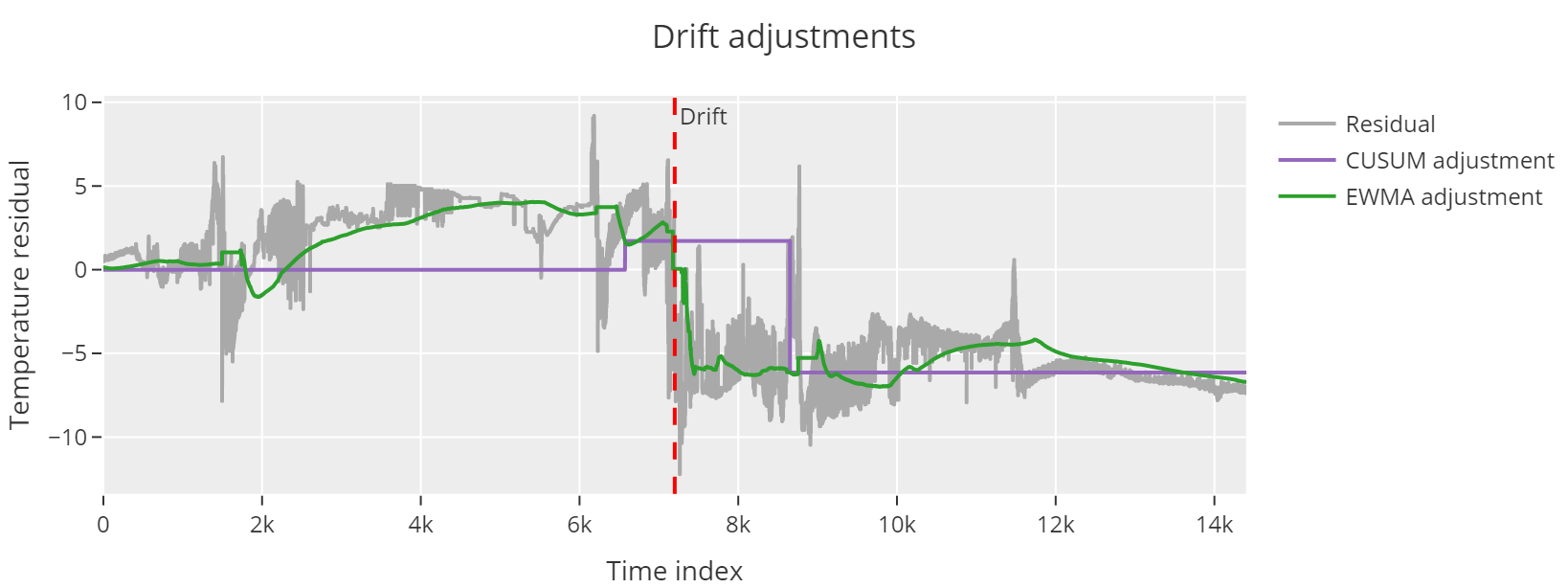}
    \includegraphics[width=0.9\textwidth]{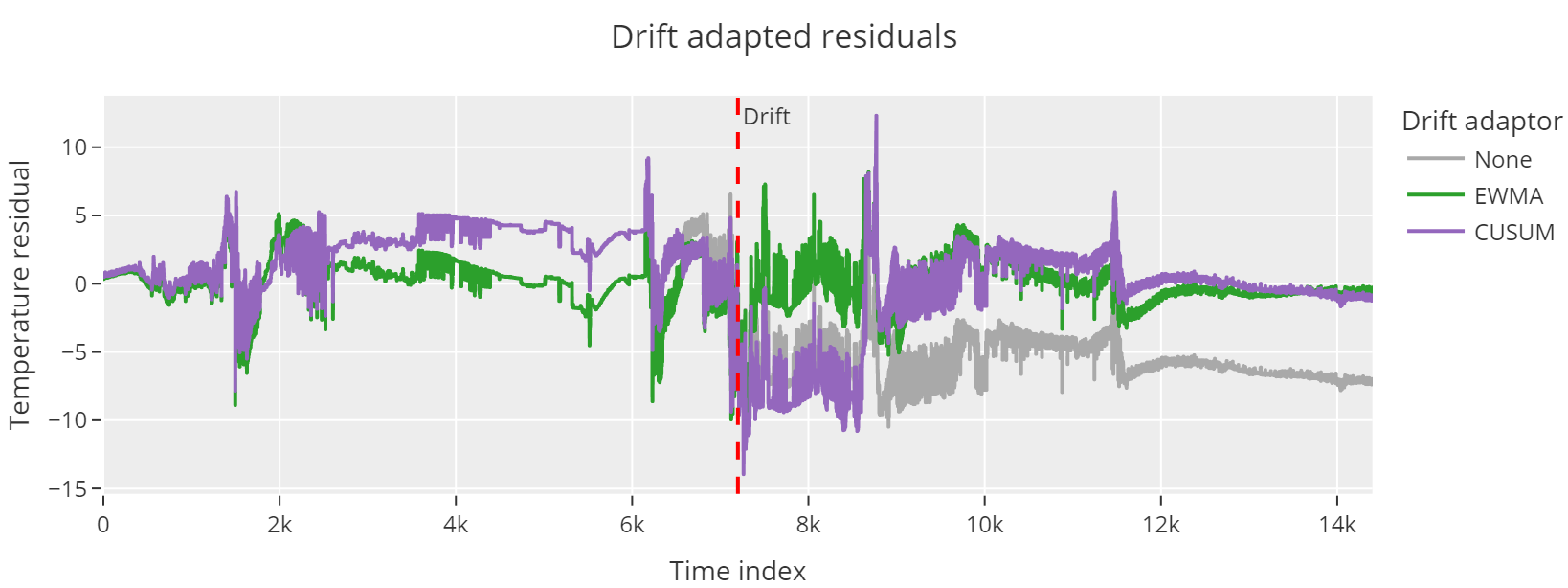}
    \caption{Drift adjustments $\hat{b}_i$ (top) and drift adjusted residuals $\tilde{e}_t = e_t - \hat{b}_t$ (bottom) for both the CUSUM and EWMA drift adaptors. Ten days of data is shown around the time of a true drift, indicated by the dashed red line. The hyperparameters of the CUSUM and the EWMA are set to the values used in the simulation experiments, described in Appendix B.}
    \label{fig:drift-adaptation}
\end{figure}

Regarding adapting to a detected drift, the CUSUM drift adaptor waits $n_\text{retrain}-l$ samples before it sets the drift adjustment $\hat{b}_t$ to the mean of the previous $n_\text{retrain}$ residuals.
In other words, if drifts are detected by $\mathcal{C}_t(l)$ at times $\mathcal{D} = \{d_i\}_{i}^{q} \subseteq \{1, \ldots, t\}$, the sequence of drift adjustments becomes
\begin{equation}
    \hat{b}_i = 
    \begin{cases} 
        0 & 1 \leq i < d_1 - l + n_\text{retrain} \\
        \bar{e}_{d_1-l+1:d_1-l+n_\text{retrain}} & d_1 - l + n_\text{retrain} \leq i < d_2 - l + n_\text{retrain} \\
        &\vdots \\
        \bar{e}_{d_q-l+1:d_q-l+n_\text{retrain}} & d_q - l + n_\text{retrain} \leq i \leq t \\
    \end{cases}
    \label{eq:cusum_adaptation}
\end{equation}
A complete summary of the CUSUM drift adaptor in a form that fits $\text{AdaptDrift}(e_t)$ in Algorithm \ref{alg:high-level-alg} is given in Algorithm \ref{alg:cusum-drift-alg}.
In addition, Figure \ref{fig:drift-adaptation} illustrates the behaviour of the method around the time of the real drift shown in Figure \ref{fig:real_drift}.
Notice the discrete jump in drift adjustment as a consequence of a detected drift $n_\text{retrain}$ samples earlier.

\begin{algorithm}[ht]
  \caption{CUSUM drift detection and adaptation}
  \label{alg:cusum-drift-alg}
  \begin{algorithmic}[1]
  \Parameters $n_\text{retrain}$, $\lambda$, $\mathcal{S}$
  \Stored History of $e_t$ and $S_{s+1:t}$ for $s \in \mathcal{S}$, drift detection times $\mathcal{D}$, initialised as an empty set, and the previous drift adjustment $\hat{b}_{t-1}$, initialised at $0$.
  \Input $e_t$
   \State $\hat{b}_t = \hat{b}_{t-1}$
   \State $S_{s+1:t} = S_{s:t-1} + e_t - e_{s}$ for $s \in \mathcal{S}$.
   \If{$t - \max(\mathcal{D}) > n_\text{retrain}$}
   \State $\mathcal{C}_{s+1:t} = S_{s+1:t} /\sqrt{t-s}$
   \State $\mathcal{C}_t = \sum_{j=1}^p \max_{s \in \mathcal{S}} |\mathcal{C}_{j, s+1:t}|$
   \If{$\mathcal{C}_t > \lambda$}
   \State $\mathcal{D} = \mathcal{D} \cup \{t\}$
   \EndIf
   \ElsIf{$t - \max(\mathcal{D}) = n_\text{retrain}$}
   \State $\hat{b}_t = \bar{e}_{t-n_\text{retrain} + 1:t}$
   \EndIf
   \MyReturn $\tilde{e}_t = e_t - \hat{b}_t$
  \end{algorithmic}
\end{algorithm}

\subsubsection{Continuous drift adaptation}  \label{sec:continuous_concept_drift}
As a continuously updated drift adaptor we suggest to use a lagged EWMA, given by
\begin{equation}
   \hat{b}_t(\alpha, l) = (1 - \alpha) \hat{b}_{t-l-1} + \alpha e_{t-l}.
\end{equation}

The lag parameter $l$'s role is the same as for the CUSUM method: To avoid masking real faults.

See Figure \ref{fig:drift-adaptation} for an illustration of how the EWMA adaptor behaves compared to the CUSUM method.
Notice the smooth adaptation compared to the discrete jumps of the CUSUM method.

\subsection{Detecting anomalous sequences of residuals}  \label{sec:anomaly_detection_method}
In this section we briefly describe the algorithm for detecting anomalies in the residuals developed by \citet{hellton_real-time_2021}, and fit it within the fault detection framework and $\text{DetectAnomaly}$ of Algorithm \ref{alg:high-level-alg}.

The mentioned algorithm is a sequential changepoint detection method for detecting consistently large positive deviations in a data sequence, originally due to \citet{lorden_sequential_2008} and \citet{liu_scalable_2017}.
It requires two hyperparameters to be set: A detection threshold $\gamma$ and a smallest relevant size of the change in mean $\rho$.
Like for the drift detector in Section \ref{sec:abrupt_concept_drift}, the top level of the algorithm consists of aggregating per-sensor scores, but this time the maximum is used as aggregator to retain sensitivity to potential hotspots forming in the motor.
Thus, the global anomaly score is given by
\begin{equation}
  G_t = \max_{j=1, \ldots p} z_{j, t},
  \label{eq:global_anomaly_score}
\end{equation}
where $z_{j, t}$ are per-sensor scores given by the adaptive recursive CUSUMs,
\begin{equation}
    z_{j, t} = \max\left(z_{j, t - 1} + \hat{\mu}_{j, t} \tilde{e}_{j, t} - \frac{1}{2}\hat{\mu}_{j, t}^2, 0\right).
    \label{eq:adaptive_cusum}
\end{equation}
Further, the mean estimates $\hat{\mu}_{j, t}$ are updated as running means from the most recent time $z_{j, t} = 0$, but bounded below at $\rho$. It is calculated as follows:
\begin{equation}
    \hat{\mu}_{j, t} =\max\left(\frac{s_{j, t}}{n_{j, t}}, \rho\right), \quad 
     s_{j, t} = \begin{cases} 
    s_{j, t - 1} + \tilde{e}_{j, t - 1}, & z_{j, t - 1} > 0, \\
    0, & z_{j, t - 1} = 0, 
   \end{cases} \label{eq:eff_cusum_mean}
\end{equation}
where $n_{j, t} = n_{j, t - 1} + 1$, if $z_{j, t - 1} > 0$, and otherwise $n_{j, t} = 0$, if $z_{j, t - 1} = 0$, and with initial values $z_{j, 0} = s_{j, 0} = \tilde{e}_{j, 0} = 0$.
When $s_{j, t} = n_{j, t} = 0$, we set $s_{j, t} / n_{j, t} = 0$. 
Finally, an alarm is raised as soon as $G_t > \gamma$.

\subsection{Tuning the change detectors} \label{sec:tuning}
To set the CUSUM drift detection threshold $\lambda$ and the anomaly detection threshold $\gamma$ we use a data driven procedure that relates the choice of threshold to a number of acceptable false detections, $q_\text{val}$.
In this way, a user of the methods sets an acceptable number of false alarms rather than the threshold directly, which is easier to interpret and reason about.

It is important that this procedure is run on a separate validation dataset different from the data used to train the prediction model.
In this way, the residuals are based on out-of-sample predictions like in a real monitoring setting.
Moreover, the validation dataset should be free of drifts and faults.

For the CUSUM drift detection score $\mathcal{C}_t$, the steps are as follows:

\begin{enumerate}
    \item Run the detector over the residuals from the validation set to obtain $\mathcal{C}_1, \ldots, \mathcal{C}_{n_\text{val}}$, where $n_\text{val}$ is the size of the validation set. 
    \item Start with $\mathcal{T} = \{1, \ldots, n_\text{val}\}$, then iteratively, for $q_\text{val}$ iterations, find $\max_{t \in \mathcal{T}} \mathcal{C}_t$ and remove a period around $\tau = \text{argmax}_{t \in \mathcal{T}} \mathcal{C}_t$ from $\mathcal{T}$.
    We choose to remove such a period by stepping to the left and right of $\tau$ until the value of $\mathcal{C}_t$ is below a threshold in both directions.
    We set the threshold to the 0.2 quantile of $\mathcal{C}_1, \ldots, \mathcal{C}_{n_\text{val}}$.
    \item The value of $\max_{t \in \mathcal{T}} \mathcal{C}_t$ from the last iteration is the tuned threshold at $q_\text{val}$ acceptable false alarms in the drift-free validation data.
\end{enumerate}

The procedure can be adapted to the anomaly detection score by exchanging $\mathcal{C}_t$ with $G_t$.

\subsection{Performance assessment}

\subsubsection{Training, validation and testing pipeline}
As our test set, we take the data from two months before the drift and onward for each ship.
If there is no drift on a ship, we use the last 25\% of samples as the test set instead. 
The remaining data is split equally between training and validation sets, where the first half is the training set, and the latter half is the validation set. 
This split results in 29.3\% training, 29.3\% validation and 41.4\% testing observations overall, but with large variations from ship to ship.

On a high level, the pipeline for training, tuning and testing is given by the following steps:
\begin{enumerate}
    \item Train the prediction model on the training set.
    \item Tune the anomaly and drift detectors on the residuals of the predictions in the validation set.
    \item Run the fault detector with drift adaptation (Algorithm \ref{alg:high-level-alg}) on the test set.
\end{enumerate}

\subsubsection{Fault simulation} \label{sec:fault_simulation}

Given that overheating faults in this class of motors are rare, we choose to simulate faults to obtain a sufficient span of different fault scenarios. The fault model is generic, but based on a real overheating event.

We use the following simplified fault model:
Each fault starts at time $u$ at some location inside the motor.
Let $\tilde{y}_t$ represent the temperature at this particular location.
From time $u$ and initial temperature $\tilde{y}_u$, the temperature increases linearly until it reaches a temperature of $145^{\circ}$C.
That is, for a slope parameter $a > 0$, $\tilde{y}_t =\tilde{y}_{u} + a(t - u)$ until $\tilde{y}_t \geq 145$.
The maximum temperature of $145^{\circ}$C is chosen as the failure temperature in these simulations. 
The time the temperature threshold is crossed is the failure time $v$.
Further we assume that only one of the sensors are affected by the latent motor fault, the fault is a hotspot and represents a worst-case scenario.
The affected sensor is sampled at random.
We model the temperature at the affected sensor $j$ by a delay $d$ as a delayed version of $\tilde{y}_t$, such that $y_{j, t + d} = \tilde{y}_t$.
The delay is sampled uniformly between $0$ and an upper limit $d_{\max}$.

In the experiments, we set the slope $a=0.62$ degrees Celsius per minute and $d_{\max}=17$ minutes.
These settings are obtained by analysing one known fault that represents a near-worst-case scenario.

\subsubsection{Performance evaluation}
A detection is recorded as a true positive if it is raised within a fault interval with start time $u$ and end time $v$, and a false positive otherwise.
A false negative or missed detection occurs if a fault interval does not contain any detections.
In addition, for each true positive detection $a_i$, we record the \textit{time to detection} as $a_i - u$ and the \textit{time to failure} as $v - a_i$.
Note that the time to failure measures how much time remains to react before the catastrophic failure occurs.

To evaluate the performance across all ships and motors, we use three standard performance measures:
Precision, recall and median time to detection.
Let $FP$, $TP$ and $FN$ to denote the total number of false positives, true positives and false negatives, respectively.
Precision is the proportion of detections that were true, given by $precision = TP / (TP + FP)$.
Recall is the proportion of true faults that were detected, given by $recall = TP / (TP + FN)$.
To summarise the times to detection, we use the median since it is more representative of a typical time to detection than the mean since the distribution of time to detection can be heavy tailed.

\subsubsection{Experimental setup} \label{sec:experimental_setup}
The purpose of the experiment is to observe how drift adaptation affects fault detection performance.
We consider three scenarios: 
Negative drift, positive drift and no drift. 
The negative drift scenario is the real, observed drift cases in the data, where all the drifts result in decreased baseline motor temperature. 
As the direction of drift influences the fault detection performance in fundamentally different ways (see the examples in Section \ref{sec:introconcept}), we have also added simulated drifts where the direction of drift has been changed from a negative to a positive direction.
That is, we have calculate the the difference in mean temperature before and after each drift, and change the sign of the difference to get the positive drift scenario.
To measure performance in the no-drift scenario, we remove any drifts present in the data.

For each of the drift scenarios, 1000 simulated overheating events are injected in the test set according to the fault simulation procedure in Section \ref{sec:fault_simulation}.
The faults are evenly distributed across motors, giving approximately 22 faults for each of the 45 motors in the dataset.
The onset times per motor are drawn uniformly from all timestamps, but restricted such that two faults are separated by a minimum of 48 hours of non-faulty observations.

For each scenario, we run the following four methods, with names as used in the figures:
\begin{itemize}
    \item CUSUM adapt: Algorithm \ref{alg:high-level-alg} with the CUSUM drift detector (Section \ref{sec:abrupt_concept_drift}) and the adaptive CUSUM anomaly detector (Section \ref{sec:anomaly_detection_method}).
    \item EWMA adapt: Algorithm \ref{alg:high-level-alg} with the EWMA drift detector (Section \ref{sec:continuous_concept_drift}) and the adaptive CUSUM anomaly detector (Section \ref{sec:anomaly_detection_method}).
    \item No adapt: Algorithm \ref{alg:high-level-alg} with no drift detector and the adaptive CUSUM anomaly detector (Section \ref{sec:anomaly_detection_method}).
    \item Threshold: Fault detection by a fixed temperature threshold. 
\end{itemize}
The parameter settings for the drift adaptation and anomaly detection methods are given in Appendix A. 

\section{Results} \label{sec:results}
The results in Table \ref{tab:scores_table_slope=1.0} shows performance of each of the proposed methods for the three different concept drift scenarios. 
The methods have been tuned such that 5 false alarms are raised in the validation dataset, giving approximately one false alarm per 11 years of operation. This corresponds to a fixed temperature threshold of $130$°C, and a detection threshold for the other methods to $\gamma = 14473$. 

\begin{table}[tb]
\caption{Performance of the considered methods in the positive, negative and no drift scenarios. The detection thresholds for all methods have been set to a value corresponding to 5 false alarms across all ships and motors in the validation data:  $\gamma = 14473$ for the EWMA, CUSUM and No adapt methods, and 130°C for the Threshold method.}
\label{tab:scores_table_slope=1.0}
\begin{tabular}{llrrrrrrr}
\toprule
Drift & Method & FP & FN & Precision & Recall & \makecell{Median time \\ to detection} & \makecell{Median time \\ to failure} \\
\midrule
Positive & CUSUM adapt & 10 & 95 & 0.988 & 0.895 & 72 & 42 \\
 & EWMA adapt & 4 & 95 & 0.995 & 0.895 & 72 & 42 \\
 & No adapt & 189 & 58 & 0.818 & 0.936 & 65 & 49 \\
 & Threshold & 1 & 0 & 0.999 & 1.000 & 98 & 15 \\
No drift & CUSUM adapt & 9 & 61 & 0.990 & 0.938 & 73 & 49 \\
 & EWMA adapt & 4 & 57 & 0.996 & 0.942 & 73 & 49 \\
 & No adapt & 13 & 66 & 0.986 & 0.933 & 73 & 49 \\
 & Threshold & 0 & 0 & 1.000 & 1.000 & 106 & 15 \\
Negative & CUSUM adapt & 9 & 30 & 0.991 & 0.970 & 73 & 56 \\
 & EWMA adapt & 4 & 30 & 0.996 & 0.970 & 73 & 56 \\
 & No adapt & 4 & 68 & 0.996 & 0.931 & 82 & 48 \\
 & Threshold & 0 & 0 & 1.000 & 1.000 & 115 & 15 \\
\bottomrule
\end{tabular}
\end{table}

\begin{figure}
    \centering
    \includegraphics[width=\textwidth]{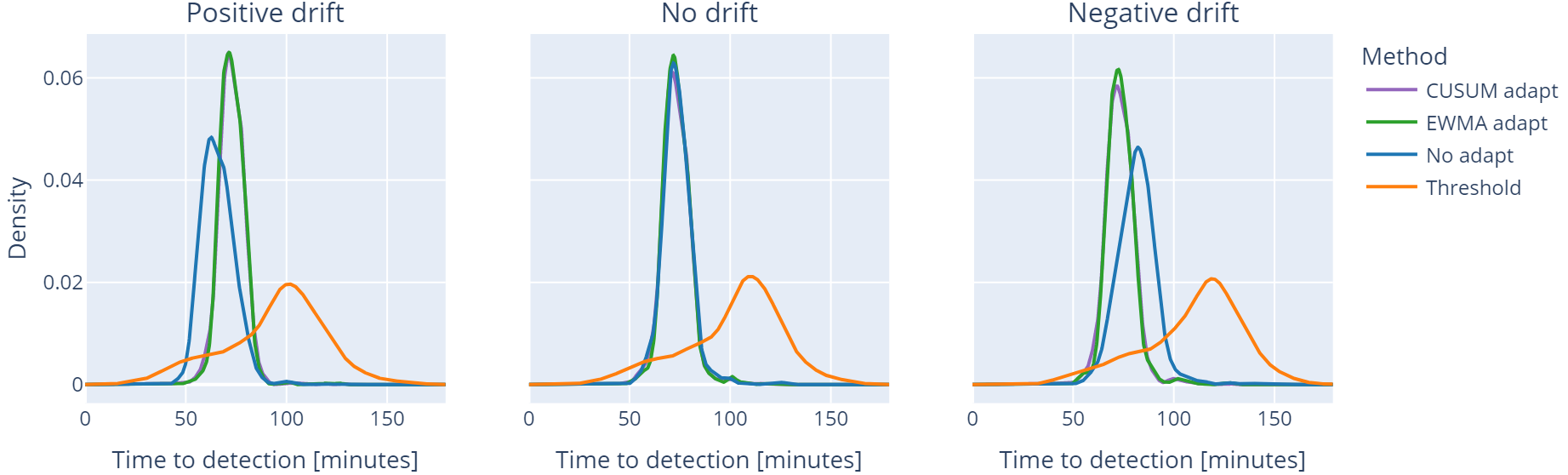}
    \caption{The distribution of time to detection for each method in the positive, no and negative drift scenarios. The distributions have been estimated by a kernel density estimator. The detection thresholds for each method correspond to five false alarms in the validation data. Observe that the residual-based CUSUM, EWMA and No adapt methods result in significantly quicker detection of faults in most cases.}
    \label{fig:ttd_density_slope1}
\end{figure}

From the table it is clear that the Threshold method performs best in terms of precision and recall, with almost perfect scores, but this comes at the cost of a longer time to detection.
By using the Threshold method, faults are typically detected only 15 minutes before catastrophic failure would occur.
The EWMA and CUSUM methods, on the other hand, typically detects the fault $26-42$ minutes earlier than the Threshold method, giving significantly more time to react.
This makes sense as the residual-based methods can detect faults much quicker when the fault starts at lower temperatures.
Figure \ref{fig:ttd_density_slope1} shows the full distributions of times to detection observed in the experiments, not just the median.
Observe that a large majority of the simulated faults are detected quicker using the EWMA and CUSUM methods compared to the Threshold method.

Another interesting result to note in Table \ref{tab:scores_table_slope=1.0} is that not doing any drift adaptation fails in the positive drift scenario, as the precision is only 0.818 in this case, compared to above approximately 0.99 for all the other methods.
The CUSUM and EWMA drift adaptors perform relatively stably across the different drift scenarios, as expected.

For results on a range of $\gamma$ values, see Appendix C.
The overall picture is the same as described here.

\section{Discussion}\label{sec:discussion}
This article presents a solution for dealing with concept drifts in fault detection systems based on monitoring residuals from a prediction model.
The solution applies an additive adjustment term to the residuals, rather than retraining the entire model.
This allows for easily understandable, quick and on-the-fly adaptation of drifts, eliminating fault monitoring downtime due to retraining.
An underlying assumption of the solution is that faults occur over shorter time-spans than concept drifts.

For an algorithm that performs adaption to concept drift, a worst case scenario is that the model adapts to a fault such that it is masked.
As presented in Section \ref{sec:continuous_concept_drift}, a safeguard is included in both the CUSUM and the EWMA method, by introducing a lag in the estimation of the drift adjustment term.
Putting additional constraints or limits on this adjustment term is a straight forward extension. 

The proposed methods are verified using simulated overheating faults, rather than actual overheating cases. On the other hand, the simulation model is based on an actual fault and is simulated as a worst-case model, where the temperature rise is affecting one sensor only.  Given the cost and criticality of the application, we need to assume that a large dataset with faults from this particular system configuration never will be available. Fault data needs to be collected from other motors and systems, or simulated. 

The results in Table \ref{tab:scores_table_slope=1.0} show that the Threshold method with a threshold of 130 has an almost perfect detection score. However, the threshold could not be set this low without detailed historical insights into the operations of the system. Given that training data is needed for the data driven fault detectors, one potential workflow could be to collect data for a given period, and then adjust the fixed threshold and put the data driven fault detector into operation. The data driven method would hence aid the fixed temperature threshold by providing early detections. 

We have concentrated on concept drifts that result in sudden changes in the mean of the residuals.
Given that the space of possible concept drifts is infinite, this is a limiting factor. 
However, the observed drifts in the motor data were all of this type, so the methods solve the problem at hand.
Dealing with more complicated drifts is left for future work, until the data demands it.

Finally, the findings of this article indicate that implementing a mechanism to adapt to concept drift is highly beneficial, if not essential, when deploying a data-driven fault detector.
This is particularly true in complex systems such as marine propulsion motors, which operate over several decades and undergo multiple scheduled maintenance operations.
Without such a mechanism, there is a risk that both short-term and long-term changes in the system, unrelated to faults, may lead to a high rate of false detections or missed faults.

\section*{Acknowledgments}
This work was supported by Norwegian Research Council centre Big Insight project 237718.
\section*{Declaration of interests}
The authors have nothing to declare.


\bibliographystyle{tfcse}
\bibliography{overheating-drift.bib}

\newpage
{\noindent \Large \textbf{Supplementary material}}

\section*{Appendix A: Algorithm reset}
To fit the adaptive CUSUM of Section \ref{sec:anomaly_detection_method} into $\text{DetectAnomaly}$ in Algorithm \ref{alg:high-level-alg}, we need a resetting mechanism upon an alarm being raised.
This is only relevant for the simulations we are running. In practice, alarms would be dealt with by e.g. shutting down the motor system.

So far, $G_t$ is only equipped to run until a single alarm is raised, while a fault detector as we have defined it, should be able to generate several alarms.
We propose the following resetting mechanism: If an alarm is raised at $a_i$, we reset the detector, and wait $R > 0$ samples before it starts running again.
We refer to $R$ as the \textit{restart delay}.
Thus, if letting $r_i = a_i + R$ be the reset time corresponding to alarm $a_i$, and $r_0 = 0$, the alarm times are given by
\begin{equation}
    a_i = \inf\left\{  t > r_{i-1} : G_t > \gamma \right\}, \quad i = 1, \ldots, m.
\end{equation}
$\text{DetectAnomaly}(\tilde{e}_t)$ is $1$ on the times $a_i$, and $0$ otherwise.

\section*{Appendix B: Drift and anomaly detector parameter settings} \label{sec:drift_parameter_settings}

The parameters of the drift and anomaly detection methods are set based on prior knowledge about the characteristics of faults and drifts.
To obtain this prior knowledge we have analysed an actual overheating fault that constitutes a near-worst-case scenario, as well as consulted with subject matter experts.
In particular, the time from onset to catastrophic failure for the analysed fault was approximately 2 hours.
In addition, for a change to count as a drift, it needs to last for at least a day.

This insight is what guides the parameter settings used throughout the performance experiments. For example, the lag parameters are conservatively set to 4 hours to be sure the drift detectors doesn't mask an overheating fault, and the smoothing-type parameters of the drift adaptors are set to values that are on the order many hours or days. The full list of parameter settings is given below.
\begin{itemize}
\item CUSUM drift adaptor: The number of retraining samples $n_\text{retrain} = 400$ (6.7 hours), lag $l = 240$ (4 hours) and candidate changepoint set $\mathcal{S}$ corresponding to the number of minutes in 1 day, 2 days, ..., 7 days. $\lambda$ is set according to the tuning procedure described in Section \ref{sec:tuning} such that $q_\text{val} = 200$ across the validation data of all motors.
\item EWMA drift adaptor: Half-life $\tau = 480$ (8 hours) and lag $l = 240$ (4 hours).
\item Anomaly detector: Minimum change size $\rho = 30$ and reset delay $R = 1440$ (24 hours).
\end{itemize}

The minimum change size parameter $\rho$ is set based on experimentation on the validation data.
Its value is guided by the fact that the adaptive CUSUM used for anomaly detection starts increasing from $0$ for temperature residuals larger than $\rho/2$.
This means that for $\rho=30$, the temperature residuals must be $15$ or larger for the adaptive CUSUM to start increasing.

\section*{Appendix C: Results for a range of detection thresholds}

\begin{figure}
    \centering
    \includegraphics[width=\textwidth]{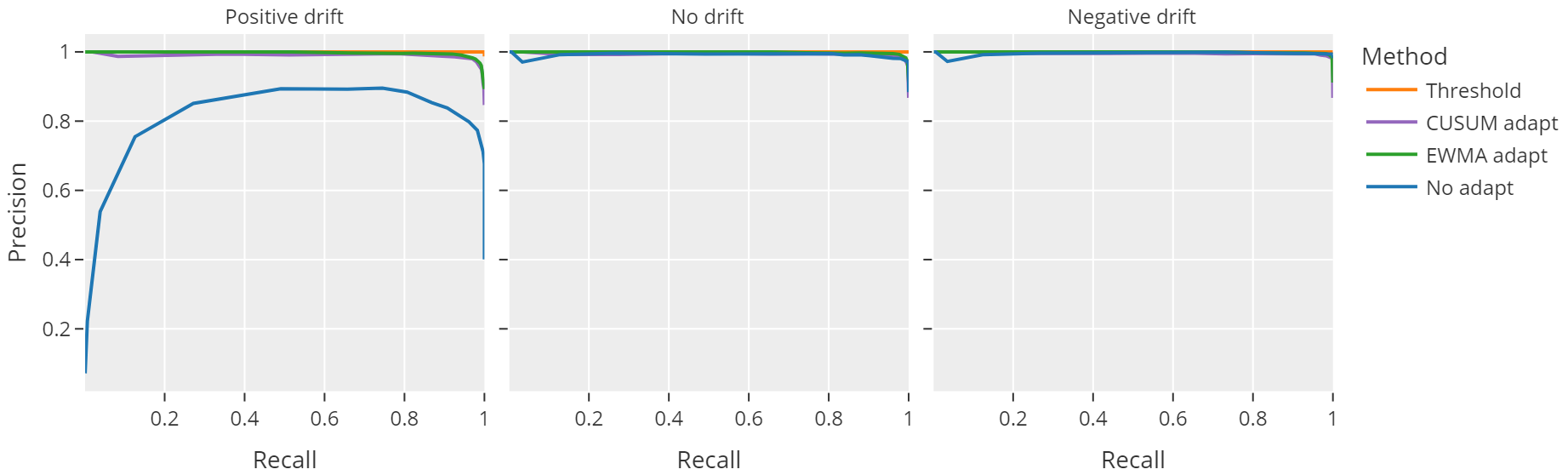}
    \caption{Precision-recall curve for each method in the positive, no and negative drift scenarios.
    Each point along the lines correspond to a different value of the anomaly detection $\gamma$ or the temperature threshold.
    Observe that all methods perform well except for the No adapt method in the positive drift case.
    }
    \label{fig:precision_vs_recall_slope1}
\end{figure}

\begin{figure}
    \centering
    \includegraphics[width=\textwidth]{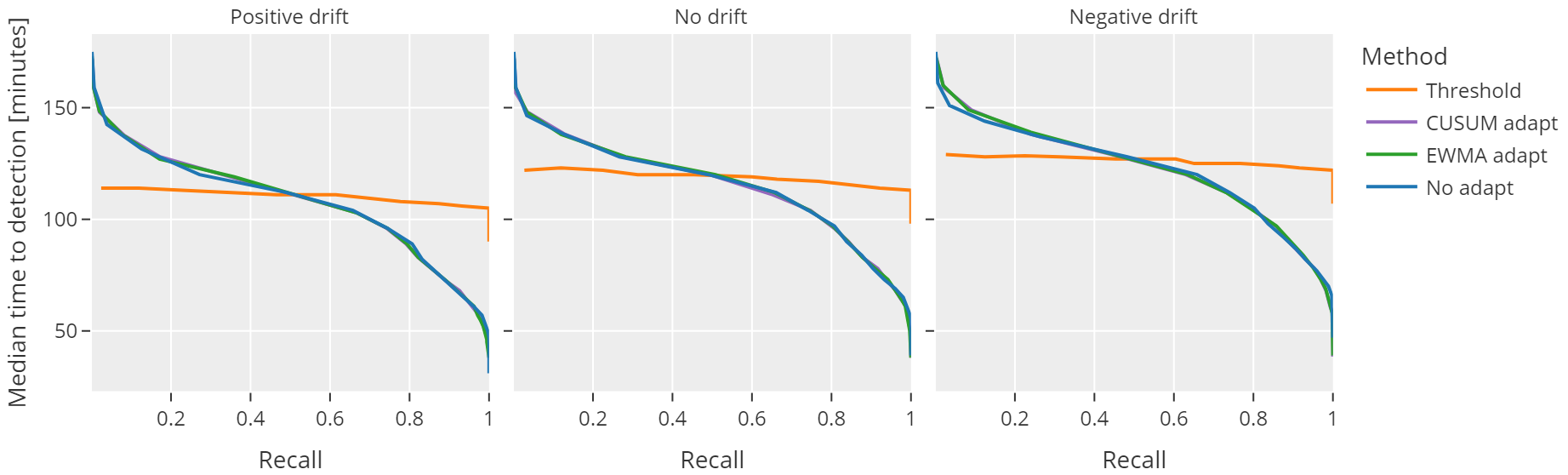}
    \caption{Median time to detection versus recall for each method in the positive, no and negative drift scenarios.
    Each point along the lines correspond to a different value of the anomaly detection $\gamma$ or the temperature threshold.
    Observe that it is almost not possible to make a trade-off between recall and time to detection for the Threshold method, as opposed to the other methods.
    }
    \label{fig:ttd_vs_recall_slope1}
\end{figure}

To further assess the performance of the methods on the test data, we also run and record performance measures for a range of different detection thresholds. This gives a more detailed picture of the relationship between detection threshold, precision, recall and time to detection.
For these experiments, we adjust the anomaly detection threshold $\gamma$ between 243 (50 false alarms in the validation data) and 258901 (0 false alarms in the validation data) and the temperature threshold between 125 to 145.

When looking at the performance for a range of detection thresholds, the overall picture is similar as for the selected threshold, but some more nuances can be observed.
Figure \ref{fig:precision_vs_recall_slope1} shows the precision-recall curves for each drift scenario and method, while Figure \ref{fig:ttd_vs_recall_slope1} shows the median time to detection-recall curves.

Again, observe in Figure \ref{fig:precision_vs_recall_slope1} that it is not possible to achieve competitive performance for the No adapt method for the positive drift scenario.

In Figure \ref{fig:ttd_vs_recall_slope1}, it is interesting to note that the time to detection has a much larger span for the CUSUM, EWMA and No adapt methods compared to the Threshold method.
This makes it possible to trade off recall with time to detection to a larger degree, depending on what is desired in a specific use case.
The Threshold method can only achieve the same time to detection performance as the residual-based methods for recall smaller than 0.5.
Also observe that the time to detection increases from positive, to no to negative drifts, but that this effect is slightly stronger for the Threshold method.


\end{document}